\documentclass[reprint,longbibliography,aps,prl]{revtex4-1}%
\usepackage[utf8]{inputenc}
\usepackage{amsmath,amssymb,amsthm,mathtools}
\usepackage{graphicx}% Include figure files
\usepackage{dcolumn}% Align table columns on decimal point
\usepackage{bm}% bold math
\usepackage{hyperref}% add hypertext capabilities
\def\be{\begin{equation}}
\def\ee{\end{equation}}
\def\ba#1\ea{\begin{align}#1\end{align}}
\def\bg#1\eg{\begin{gather}#1\end{gather}}
\def\bm#1\em{\begin{multline}#1\end{multline}}
\def\bmd#1\emd{\begin{multlined}#1\end{multlined}}

\def\se{\section}

\def\nn{\nonumber}

\def\({\left(}
\def\){\right)}
\def\[{\left[}
\def\]{\right]}
\def\<{\langle}
\def\>{\rangle}

\def\Tr{\operatorname{Tr}}

\begin{document}

%\preprint{      }
\title{Information Theoretic Inequalities as Bounds in Superconformal Field Theory}
\author{Yang Zhou}
\affiliation{School of Physics and Astronomy, Tel-Aviv University, Ramat-Aviv 69978, Israel}
\email{yangzhou@post.tau.ac.il}
\date{\today}

\begin{abstract}
An information theoretic approach to bounds in superconformal field theories is proposed. It is proved that the supersymmetric R\'enyi entropy $\bar S_\alpha$ is a monotonically decreasing function of $\alpha$ and $(\alpha-1)\bar S_\alpha$ is a concave function of $\alpha$. Under the assumption that the thermal entropy associated with the ``replica trick'' time circle is bounded from below by the charge at $\alpha\to\infty$, it is further proved that both ${\alpha-1\over \alpha}\bar S_\alpha$ and $(\alpha-1)\bar S_\alpha$ monotonically increase as functions of $\alpha$. Because $\bar S_\alpha$ enjoys universal relations with the Weyl anomaly coefficients in even-dimensional superconformal field theories, one therefore obtains a set of bounds on these coefficients by imposing the inequalities of $\bar S_\alpha$. Some of the bounds coincide with Hofman-Maldacena bounds and the others are new. We also check the inequalities for examples in odd-dimensions.

\end{abstract}

\maketitle

%%%%%%%%%%%%%%%%%%%%%%%%%%%%%%%%%%%%%%%%%%%%%%%%%%
\se{Introduction}
Quantum information theoretic ideas, such as quantum entanglement, have recently played significant roles in condensed matter physics~\cite{Vidal:2002rm,Kitaev:2005dm,Levin:2006}, particle physics~\cite{Casini:2004bw,Casini:2006es,Casini:2012ei,Casini:2008cr} and string theory~\cite{Ryu:2006bv}. To characterize the entanglement in states of a quantum mechanical system, one often bipartitions the system and computes the entanglement entropy, $S_{\text{EE}}$. Another interesting measure is the R\'enyi entropy, $S_\alpha$, which is a one parameter generalization of entanglement entropy and provides additional information about the entanglement structure for the same bipartition and returns to $S_{\text{EE}}$ in the limit $\alpha\to 1$. $\alpha$ is called its order. In quantum field theory (QFT), one defines the entanglement entropy associated with a global state and a geometric region $A$ by tracing over the field variables outside $A$, creating a reduced density matrix $\rho_A$ and then evaluating $S_{\text{EE}}$~\cite{Callan:1994py}. While $S_{\text{EE}}$ (or $S_\alpha$) generally includes UV divergences in QFT, its universal part contains important physical information, such as central charges characterizing degrees of freedom~\cite{Solodukhin:2008dh,Myers:2010tj,Casini:2011kv,Jafferis:2011zi,Liu:2012eea}. In many aspects, these universal terms are the counterparts of quantum-mechanical entropies, which satisfy a set of inequalities inspired from information theory. One natural question is: what are the QFT counterparts of these entropy inequalities and what are their roles? One inequality of $S_{\text{EE}}$ called strong sub-additivity plays significant roles in constructing monotonically decreasing $c$-functions along RG flows, such as the two-dimensional entropic $c$-function~\cite{Casini:2004bw,Casini:2006es} and the three-dimensional $F$-function~\cite{Casini:2012ei,Liu:2012eea}. Other applications of information theoretic inequalities include refining Bekenstein bound~\cite{Casini:2004bw}, deriving the integrated null energy condition~\cite{Faulkner:2016mzt} and deriving gravitational positive energy conditions~\cite{Lashkari:2016idm}.

In this letter we concern the R\'enyi entropy inequalities related to its order $\alpha$, which were proven in information theory~\cite{BeckSchlogl} and still hold in quantum mechanics~\cite{quantumP}. One therefore expects that these inequalities also play significant roles in QFT~\cite{HR}. However, the exact results of R\'enyi entropy are very rare in QFT (except for $2d$ conformal field theories)~\cite{Calabrese:2004eu,Calabrese:2009qy,Casini:2010kt,Klebanov:2011uf,Fursaev:2012mp,Dowker:2012rp,Lee:2014zaa}. We therefore focus on a subset of field theories, supersymmetric ones with a conserved R-symmetry. By twisting the ordinary R\'enyi entropy to be supersymmetric~\cite{NY}, $S_\alpha\to\bar S_\alpha$, we are able to obtain exact results at any coupling. For even-dimensional superconformal field theories (SCFTs), the supersymmetric R\'enyi entropy $\bar S_\alpha$ enjoys universal relations with the Weyl anomaly coefficients. These relations are independent of the specific theory and therefore can be used to bound the space of SCFTs. That is, imposing $\bar S_\alpha$'s inequalities to these relations gives a set of bounds on the Weyl anomaly coefficients. The key step in this derivation is to find the inequalities satisfied by $\bar S_\alpha$, which is the main topic of this letter. The idea is that, $\bar S_\alpha$ can be expressed as the R\'enyi divergence of the energy distribution from the R-charge distribution. By studying the $\alpha$-dependence of the R\'enyi divergence, one can get the inequalities satisfied by $\bar S_\alpha$. It is proved along this way that $\bar S_\alpha$ monotonically decreases as a function of $\alpha$ and $(\alpha-1)\bar S_\alpha$ is a concave function of $\alpha$. On the other hand, $\bar S_\alpha$ of CFTs associated with a spherical entangling surface is related to other physical quantities such as thermal entropy $S$, energy $E$ and charge $Q$ defined on the hyperbolic space $\Bbb{S}^1_\alpha\times\Bbb{H}^{d-1}$~\cite{Huang:2014gca}. Under the assumption that the thermal entropy is bounded from below by the charge at $\alpha\to\infty$, it is further proved that both ${\alpha-1\over \alpha}\bar S_\alpha$ and $(\alpha-1)\bar S_\alpha$ monotonically increase as functions of $\alpha$.

We will start by introducing R\'enyi divergence in information theory and studying its behavior as a function of $\alpha$, which will be used for the later proof of the supersymmetric R\'enyi entropy inequalities.    
Then the applications of these inequalities in even dimensions will be discussed and the validity of them will be checked for some odd-dimensional examples. A holographic derivation of the bound $S\geq2\pi Q$ will be given in the appendix.

%%%%%%%%%%%%%%%%%%%%%%%%%%%%%%%%%%%%%%%%%%%%%%%%%%
\section{R\'enyi Divergence}\label{Rdiv}
In information theory, R\'enyi divergence is related to R\'enyi entropy much like Kullback-Leibler divergence (relative entropy) is related to Shannon entropy. For a probability distribution $P=(p_1,\dots, p_n)$, which satisfies $p_i\geq 0$ and $\sum_{i=1}^n p_i = 1$, the Shannon entropy is given by
\be\label{ShannonE}
H(P)=-\sum_{i=1}^n p_i\log p_i\ ,
\ee and the R\'enyi entropy is given by ($\alpha>0$)
\be\label{renyiE}
H_\alpha(P)={1\over 1-\alpha}\log\sum_{i=1}^n p_i^\alpha\ ,
\ee which reduces to the Shannon entropy (\ref{ShannonE}) in the limit $\alpha\to 1$ and can be considered as the $\alpha$-extension of the Shannon entropy.
Let $Q$ be another probability distribution, $Q=(q_1,\dots, q_n)$. The relative entropy between $P$ and $Q$ is given by
\be\label{relativeE}
D(P||Q) = \sum_{i=1}^n p_i\log{p_i\over q_i}\ ,
\ee which can be proven to be nonnegative for two normalized distributions $P$ and $Q$. Notice that the relative entropy is regular only if $q_i=0$ implies $p_i=0$ for all $i$, in another word $P$ is absolutely continuous with respect to $Q$, $P\ll Q$. In our later set up in QFT, $P$ and $Q$ will be identified as energy distribution ($\propto e^{-K}$) and charge distribution ($\propto e^{-\hat Q^\prime}$), respectively. Therefore $P\ll Q$ is guaranteed by the Bogomol'nyi-Prasad-Sommerfield bound. The $\alpha$-extension of the relative entropy (\ref{relativeE}) is the R\'enyi divergence, 
\be\label{renyiD}
D_\alpha(P||Q)={1\over \alpha-1}\log\sum_{i=1}^n p_i^\alpha q_i^{1-\alpha}\ ,
\ee
which was introduced by R\'enyi as a measure of information that satisfies almost the same axioms as the relative entropy~\cite{Renyi}. In particular, the R\'enyi divergence reduces to the relative entropy in the limit $\alpha\to 1$. On the other hand,
one may consider the R\'enyi divergence (\ref{renyiD}) as a deformation of the R\'enyi entropy (\ref{renyiE}). Indeed, the R\'enyi entropy can be expressed in terms the R\'enyi divergence of $P$ from the uniform distribution $U=(1/n,\dots, 1/n)$:
\be\label{connection}
H_\alpha(P)=H_\alpha(U)-D_\alpha(P||U)=\log n-D_\alpha(P||U)\ .
\ee

Let us study the $\alpha$-dependence of the R\'enyi divergence. In order to understand that, one may first look at the $\alpha$-related R\'enyi entropy inequalities:
\ba
\partial_\alpha H_\alpha &\leq 0\ ,\label{ineq1}\\
\partial_\alpha \left({\alpha-1\over \alpha}H_\alpha\right) &\geq 0\ ,\label{ineq2}\\
\partial_\alpha \left((\alpha-1)H_\alpha\right) &\geq 0\ ,\label{ineq3}\\
\partial_\alpha^2 \left((\alpha-1)H_\alpha\right) &\leq 0\ .\label{ineq4}
\ea
For the proof of these inequalities, we refer to~\cite{BeckSchlogl}. One natural question is: Are there similar inequalities like (\ref{ineq1})-(\ref{ineq4}) for the R\'enyi divergence $D_\alpha$?
We are now going to prove the following inequalities:
\ba
\partial_\alpha D_\alpha &\geq 0\ ,~\alpha>0\label{ineq5}\\
\partial_\alpha \left({\alpha-1\over \alpha} D_\alpha\right) &\geq 0\ ,~\alpha>0\label{ineq6}\\
\partial_\alpha \left((\alpha-1) D_\alpha\right) &\geq 0\ ,~\alpha\geq 1\label{ineq7}\\
\partial_\alpha^2 \left((\alpha-1) D_\alpha\right) &\geq 0\ ,~\alpha>0\ .\label{ineq8}
\ea
Among these four inequalities, (\ref{ineq5}) and (\ref{ineq8}) have been proven by Tim van Erven and Peter Harremo\"{e}s in~\cite{RDivergence}. We will prove the other two (\ref{ineq6}) and (\ref{ineq7}) and give an alternative proof of (\ref{ineq8}). Below we also include the proof of (\ref{ineq5}) by Tim van Erven and Peter Harremo\"{e}s for completeness.
\subsection{Monotonicity of $D_\alpha$}
Now we prove $\partial_\alpha D_\alpha \geq 0$. Let $\alpha<\beta$ be positive real numbers ($\alpha,\beta\neq 1$). Then for $x\geq 0$, the function $f(x)=x^{\alpha-1\over \beta-1}$ is strictly convex if $\alpha<1$ and strictly concave if $\alpha>1$. Therefore by Jensen's inequality
\ba
D_\alpha &= {1\over \alpha -1} \log\sum_{i=1}^n\left({p_i\over q_i}\right)^{(\beta-1){\alpha-1\over\beta-1}} p_i\\
&\leq{1\over \beta-1}\log\sum_{i=1}^n\left({p_i\over q_i}\right)^{\beta-1} p_i\label{firstJensen}\\
&= D_\beta\ ,
\ea Notice that the normalization condition $\sum_i q_i =1$ is not necessary for the proof of $\partial_\alpha D_\alpha \geq 0$. Jensen's inequality states that, if $f(x)$ is a convex function of $x$, then
\be
\text{E}[f(x)]\geq f(\text{E}[x])\ ,
\ee where $\text{E}[X]$ means taking the average of variable $X$ under a normalized probability distribution. The inequality is reversed if $f(x)$ is concave.

\subsection{Monotonicity of ${\alpha-1\over \alpha} D_\alpha$}
Now we prove $\partial_\alpha\left({\alpha-1\over \alpha} D_\alpha\right)\geq0$. Let $\alpha<\beta$ be positive real numbers. Then for $x\geq 0$, the function $f(x)=x^{\alpha\over\beta}$ is strictly concave. Therefore
\ba
{\alpha-1\over\alpha}D_\alpha %&= {1\over \alpha}\log\sum_{i=1}^n p_i^\alpha q_i^{1-\alpha}\\
&={1\over\alpha}\log\sum_{i=1}^n \left({p_i\over q_i}\right)^{\beta{\alpha\over\beta}} q_i\\
&\leq{1\over\beta}\log\sum_{i=1}^n \left({p_i\over q_i}\right)^{\beta} q_i\\
&= {\beta-1\over\beta}D_\beta\ ,
\ea where we have used Jensen's inequality again in the second step. Notice that the normalization condition $\sum_i q_i=1$ now is essential in this proof.

\subsection{Monotonicity of $(\alpha-1) D_\alpha$}
Now we prove $\partial_\alpha \left((\alpha-1) D_\alpha\right)\geq 0$ for $\alpha\in [1,\infty)$. Given that $\partial_\alpha^2 \left((\alpha-1) D_\alpha\right) \geq 0$, which will be proven in the following subsection, we only need to prove
\be
\partial_\alpha \left((\alpha-1) D_\alpha\right)|_{\alpha \to 1} \geq 0\ .
\ee
This can be shown as follows
\ba
\partial_\alpha \left((\alpha-1) D_\alpha\right) &= \partial_\alpha \left(\log\sum_{i=1}^n p_i^\alpha q_i^{1-\alpha}\right) \\
&= {\sum_{i=1}^n\left(p_i\over q_i\right)^\alpha q_i\log{p_i\over q_i}\over \sum_{i=1}^n \left({p_i\over q_i}\right)^\alpha q_i}\ .\label{firstD}
\ea
The $\alpha\to 1$ limit of the above formula can be evaluated
\be
\partial_\alpha \left((\alpha-1) D_\alpha\right)|_{\alpha \to 1} = D(P||Q)\geq 0\ .
\ee
In the last step we have used the non-negativity of the relative entropy, whose proof requires the normalization conditions for both $P$ and $Q$.
\subsection{Convexity of $(\alpha-1) D_\alpha$}
Now we prove $\partial_\alpha^2 \left((\alpha-1) D_\alpha\right) \geq 0$. We take one more derivative of (\ref{firstD}) with respect to $\alpha$
\ba
\partial_\alpha^2 \left((\alpha-1) D_\alpha\right) &= {\sum_{i=1}^n\left(p_i\over q_i\right)^\alpha q_i\left(\log{p_i\over q_i}\right)^2\over \sum_{i=1}^n \left({p_i\over q_i}\right)^\alpha q_i}\nonumber\\ &- {\left[\sum_{i=1}^n\left(p_i\over q_i\right)^\alpha q_i\log{p_i\over q_i}\right]^2\over \left[\sum_{i=1}^n \left({p_i\over q_i}\right)^\alpha q_i\right]^2}\ .\label{secondD}
\ea
Define a new distribution
\be
\bar\rho_i := {\left({p_i\over q_i}\right)^\alpha q_i\over \sum_{i=1}^n \left({p_i\over q_i}\right)^\alpha q_i}\ ,
\ee which automatically satisfies the normalization $\sum_i \bar\rho_i =1$, one can rewrite (\ref{secondD}) as
\ba
\partial_\alpha^2 \left((\alpha-1) D_\alpha\right) &= \sum_{i=1}^n\bar\rho_i\left(\log{p_i\over q_i}\right)^2 - \left(\sum_{i=1}^n\bar\rho_i\log{p_i\over q_i}\right)^2\nonumber\\
&\geq 0\ ,
\ea where the last step follows from the convexity of the function $f(x)=x^2$ and Jensen's inequality.

%%%%%%%%%%%%%%%%%%%%%%%%%%%%%%%%%%%%%%%%%%%%%%%%%%
\se{Inequalities of supersymmetric R\'enyi entropy}
In QFTs in flat space, the R\'enyi entropy can be used to measure the entanglement spectrum between two regions $A$ and $\bar A$ separated by the entangling surface $\Sigma$. For a state characterized by a density matrix $\rho_0$ on a spatial slice consisting of $A$ and $\bar A$, one can define the R\'enyi entropy for $A$ using the reduced density matrix $\rho_A=\Tr_{\bar A}\rho_0$,
\be\label{Frenyi}
S_\alpha = {1\over 1-\alpha}\log\Tr\rho_A^{\,\alpha}\ ,
\ee
where $\alpha>0$.  
$S_\alpha$ in (\ref{Frenyi}) is the field theory analogy of $H_\alpha$ in (\ref{renyiE}) in information theory. Notice that the previous definition now has been generalized to infinite dimensional spaces by replacing the probabilities by the reduced density matrix and the sum by a trace. In Euclidean QFT, a state is characterized by a path integral with certain boundary conditions. Therefore (\ref{Frenyi}) can be expressed in terms of path integrals on a Euclidean spacetime with a conical singularity.

We focus on CFTs in $\Bbb{R}^{1,d-1}$, the R\'enyi entropy (\ref{Frenyi}) associated with a spherical entangling surface ($\Sigma=\Bbb{S}^{d-2}$) can be computed by conformally mapping the Euclidean conic space to a hyperbolic space $\Bbb{S}^1_\alpha\times\Bbb{H}^{d-1}$, where the previous density matrix $\rho_A$ now becomes $\rho\propto e^{-2\pi H}$ by a unitary transformation and $H$ is the Hamiltonian quantized on $\Bbb{H}^{d-1}$~\cite{Casini:2011kv}. In this case, $S_\alpha$ can be written as
\be
S_\alpha = {1\over 1-\alpha}\log\Tr \rho^\alpha = {1\over 1-\alpha}\log{\Tr e^{-2\pi\alpha H}\over \left(\Tr e^{-2\pi H}\right)^\alpha}\ ,\label{renyiHyp}
\ee where we have considered the normalization $\Tr\rho=1$.

We are particularly interested in supersymmetric theories with a conserved $U(1)$ R-symmetry because the computation of $S_\alpha$ is very challenging for interacting theories. We also restrict ourselves to the spherical entangling surface without considering the shape dependence. In the viewpoint of rigid supersymmetry, the spacetime with a conical singularity breaks all the supersymmetries. Equivalently, the space $\Bbb{S}^1_{\alpha}\times\Bbb{H}^{d-1}$ for $\alpha\neq 1$ does not preserve any supersymmetry. To proceed further, we twist the R\'enyi entropy (\ref{renyiHyp}) into a supersymmetric one by turning on a background R-symmetry gauge field along $\Bbb{S}_\alpha^1$. This twisting has been first studied in three dimensions~\cite{NY,Huang:2014gca} and then extended to other dimensions~\cite{HZ,Crossley:2014oea,5dsre,Alday:2014fsa,Hama:2014iea,Zhou:2015cpa,Nian:2015xky,Zhou:2015kaj,Mori:2015bro,Giveon:2015cgs}. The supersymmetric twist can be written as~\cite{testtrace}
\ba
\bar S_\alpha &= {1\over 1-\alpha}\log{\Tr e^{-2\pi\alpha (H-\mu \hat Q)}\over \left(\Tr e^{-2\pi H}\right)^\alpha}\ ,\\
&= {1\over 1-\alpha}\log{\Tr e^{-2\pi(\alpha H- \hat Q(\alpha-1))}\over \left(\Tr e^{-2\pi H}\right)^\alpha}\ ,\label{sRenyi}
\ea where the chemical potential corresponding to the conserved $U(1)$ R-symmetry takes the value
\be
\mu ={\alpha-1\over\alpha}\label{muv}
\ee as required by the Killing spinor equations. Notice that we choose the convention such that the preserved Killing spinors' R-charge $r=1/2$ in general $d$-dimensions.  For details on how to determine $\mu$ by solving Killing spinor equations on conic space in various dimensions, $d=2,3,4,5,6$, we refer to~\cite{Mori:2015bro}~\cite{Huang:2014gca}~\cite{HZ}~\cite{Hama:2014iea}~\cite{Nian:2015xky}, respectively. Obviously $\bar S_\alpha$ returns to $S_{\text{EE}}$ at $\alpha\to 1$. By unitarily transforming the effective density matrix in (\ref{sRenyi}),
one can rewrite $\bar S_\alpha$ in flat space by replacing $H$ with the modular Hamiltonian $K$ and replacing $\hat Q$ with the conserved R-charge $\hat Q^\prime$ defined in the subregion $A$. Note that the trace now is taken over the Hilbert space of the subregion $A$ as (\ref{Frenyi}). One can observe a connection between the supersymmetric R\'enyi entropy and the R\'enyi divergence $D_\alpha$ in (\ref{renyiD}). That is, by identifying
\be\label{defdistribution}
\rho_A = {e^{-2\pi K}\over \Tr e^{-2\pi K}}\ ,\quad \sigma_A= e^{-2\pi \hat Q^\prime}\ ,
\ee
one can express $\bar S_\alpha$ in terms of the R\'enyi divergence of $\rho_A$ from $\sigma_A$~\cite{renyiRelative}
\ba
\bar S_\alpha &= {1\over 1-\alpha}\log\Tr\,\rho_A^\alpha \sigma_A^{1-\alpha}\ ,\\
&=-D_\alpha(\rho_A||\sigma_A)\ .\label{srenyiD}
\ea
Notice that to make the identification (\ref{srenyiD}) we temporarily abandon the normalization condition for $\sigma_A$. Also notice that in our case $[\rho_A,\sigma_A]=0$ because of $[K,\hat Q^\prime]=0$, therefore we do not distinguish R\'enyi divergence and quantum R\'enyi divergence. As one can see from the previous section, the normalization condition for the second distribution is not necessary for the proof of (\ref{ineq5}) and (\ref{ineq8}). Therefore, the following two inequalities follow directly by replacing $D_\alpha$ by $-\bar S_\alpha$ in (\ref{ineq5}) and (\ref{ineq8}),
\ba
\partial_\alpha \bar S_\alpha &\leq 0\ ,\label{ineq9}\\
\partial_\alpha^2 \left((\alpha-1)\bar S_\alpha\right) &\leq 0\ .\label{ineq10}
\ea
This proves the monotonicity of $\bar S_\alpha$ and the concavity of $(\alpha-1)\bar S_\alpha$. One can think that they are the analogies of the properties (\ref{ineq1})(\ref{ineq4}) of the R\'enyi entropy. 

Now we study the other two analogies of (\ref{ineq2})(\ref{ineq3}) for $\bar S_\alpha$:
\ba
\partial_\alpha \left({\alpha-1\over \alpha}\bar S_\alpha\right) &\geq 0\ ,\label{ineq11}\\
\partial_\alpha \left((\alpha-1)\bar S_\alpha\right) &\geq 0\ .\label{ineq12}
\ea
As one can see, they can not be deduced from (\ref{ineq6}) or (\ref{ineq7}), because the normalization condition of $\sigma$ now is crucial.

We instead give a physical proof of (\ref{ineq11}) and (\ref{ineq12}) by following the way in~\cite{HR}. We begin with the supersymmetric partition function $Z$ on $\Bbb{S}_\alpha^1\times\Bbb{H}^{d-1}$ with a $U(1)$ R-symmetry chemical potential. We work in grand canonical ensemble,
\be
Z[\beta, \mu]=\text{Tr}\left[e^{-\beta(H-\mu\hat Q)}\right]\ ,\label{Z}
\ee where the inverse temperature $\beta$ and the chemical potential $\mu$ are background parameters, $\beta = 2\pi \alpha\,,\mu = {\alpha-1\over \alpha}$. Define $I:=-\log Z$, the state variables can be worked out from (\ref{Z}),
\ba
E &= \left({\partial I\over \partial\beta}\right)_\mu - {\mu\over\beta}\left({\partial I\over \partial\mu}\right)_\beta\ ,\label{Evariable}\\
S &= \beta \left({\partial I\over \partial\beta}\right)_\mu - I\ ,\\
Q &= -{1\over\beta}\left({\partial I\over \partial\mu}\right)_\beta\ .\label{Qvariable}
\ea
Therefore we get the energy expectation value $E=\text{Tr}(e^{-\beta(H-\mu \hat Q)}\,H) / \text{Tr} (e^{-\beta(H-\mu \hat Q)})$ by (\ref{Evariable}) and the charge expectation value $Q=\text{Tr} (e^{-\beta(H-\mu \hat Q)}\,\hat Q)/\text{Tr} (e^{-\beta(H-\mu \hat Q)})$ by (\ref{Qvariable}).
The thermal entropy $S$ is given by
\be\label{thS}
S=\beta (E-\mu Q)- I\ .
\ee
In the presence of supersymmetry, both the inverse temperature $\beta$ and the chemical potential $\mu$ are functions of a single variable $\alpha$ and therefore $I$ is considered as 
\be I_\alpha:=I[\beta(\alpha),\mu(\alpha)]\ .
\ee The supersymmetric R\'enyi entropy is defined as
\be\label{srePh}
\bar S_\alpha = {\alpha\over 1-\alpha}\left( I_1- {I_\alpha\over\alpha}\right) = {\alpha\over 1-\alpha}\int_\alpha^1\partial_{\alpha^\prime} \left({I_{\alpha^\prime}\over \alpha^\prime}\right)\ .
\ee
From this expression one can write
\ba
\partial_\alpha \left({\alpha-1\over \alpha}\bar S_\alpha\right) &= \partial_\alpha \left({I_\alpha\over \alpha}\right)\label{bridge}\\
&= {\beta (E-Q)-I_\alpha\over \alpha^2}\ .
\ea When $\hat Q$ vanishes, the numerator is exactly the thermal entropy (\ref{thS}), which was assumed to be positive in~\cite{HR} to prove the inequality (\ref{ineq2}) of the R\'enyi entropy. In the presence of supersymmetry, one may extend the positive thermal entropy condition to be
\be
S\geq 2\pi Q\ .\label{SboundQ}
\ee That is, the thermal entropy is bounded from below by the charge. While a general field theory argument for this bound is still lacking, we give a holographic derivation for CFTs having gravity duals in the appendix. The holographic derivation shows that this bound comes from the causality in gravitational physics. Then
\be\label{SboundQ1}
 \beta(E-Q)-I_\alpha=S-2\pi Q \geq 0\ ,
\ee
which ensures that
\be
\partial_\alpha \left({\alpha-1\over \alpha}\bar S_\alpha\right) \geq 0\ .
\ee In fact, one can rewrite the proved inequality (\ref{ineq10}) as $\partial_\alpha(E-Q)\leq 0$. Notice that the first $\alpha$-derivative of (\ref{SboundQ1}), $\partial_\alpha(\beta (E-Q)-I_\alpha)=\beta\partial_\alpha(E-Q)\leq 0$, one can prove (\ref{SboundQ1}) and therefore (\ref{ineq11}) with the minimal assumption $(S-2\pi Q)_{\alpha\to\infty}\geq 0$.

We are left to prove (\ref{ineq12}). Given the non-positivity of the second derivative of $(\alpha-1)\bar S_\alpha$, (\ref{ineq10}), the only thing we have to show is that
\be
\partial_\alpha \left[(\alpha-1)\bar S_\alpha\right] \bigg|_\infty \geq 0\ .
\ee
By using the last expression in (\ref{srePh}), we have
\ba
\partial_\alpha \left[(\alpha-1)\bar S_\alpha\right] \bigg |_\infty &= \left[\int_1^\alpha \partial_{\alpha^\prime} \left({I_{\alpha\prime}\over \alpha^\prime}\right) + \alpha\,\partial_\alpha \left({I_\alpha\over \alpha}\right)\right]_\infty\\
&\geq 0\ ,
\ea where the last step follows from the positivity of $\partial_\alpha \left({I_\alpha\over \alpha}\right)$ (\ref{bridge}).

In a summary we have shown that the inequalities (\ref{ineq9})-(\ref{ineq12}) hold for supersymmetric R\'enyi entropy under the assumption that the thermal entropy is bounded from below by the charge at $\alpha\to\infty$.
One can also express (\ref{ineq9}) in terms of the thermal entropy and the energy:
\be
{S-2\pi E+I_1\over (\alpha-1)^2}\leq 0\ ,
\ee which is equivalent to ($\alpha\neq 1$)
\be
\Delta S\leq 2\pi\Delta E\ ,\label{Bbound}
\ee
where $\Delta S:=S-S_{\alpha=1}$ and $\Delta E:=E-E_{\alpha=1}$. (\ref{Bbound}) is the Bekenstein bound under the deformation parametrized by $\delta=\alpha-1$ in the spirit of~\cite{Casini:2008cr}. This bound is independent of the charge therefore it can also be derived from the ordinary R\'enyi entropy property. One may also write (\ref{ineq12}) equivalently as $2\pi(E-Q)-I_1\geq 0$.

%%%%%%%%%%%%%%%%%%%%%%%%%%%%%%%%%%%%%%%%%%%%%%%%%%
\se{Applications}
Now we discuss the applications of the inequalities (\ref{ineq9})-(\ref{ineq12}). Our main concern is a spherical entangling surface in CFTs in flat space, the universal part of R\'enyi entropy (or supersymmetric) is invariant on $\Bbb{H}^{d-p}\times\Bbb{S}_\alpha^p$ for different integer $p$, where $\alpha$ denotes a conical singularity and $1\leq p\leq d$, since these geometries are related by Weyl transformations~\cite{Casini:2011kv,Zhou:2015kaj}. We mainly focus on $\Bbb{S}^1_\alpha\times\Bbb{H}^{d-1}$ but it is equivalent to working on other geometries such as conic sphere $\Bbb{S}_\alpha^d$. In order to avoid a sign ambiguity coming from the regularization of the volume $V_{d-1}$ of the hyperbolic space $\Bbb{H}^{d-1}$, we instead consider $s_\alpha := \bar S_\alpha/V_{d-1}$ as the true quantity in applying the inequalities (\ref{ineq9})-(\ref{ineq12}).

\subsection{Even-dimensional SCFTs}
{\it $d=2, {\cal N}=(2,2)$}\, SCFT~~~For these theories, $\bar S_\alpha$ has been computed from the partition function on branched two sphere~\cite{Mori:2015bro} or the correlation function of twisted fields~\cite{Giveon:2015cgs}. $\bar S_\alpha$ is independent of $\alpha$ and coincides with the entanglement entropy, whose log term is ${c\over 3}\log{R\over \epsilon}$ where $c$ is the $2d$ central charge and $R$ is the length of a single interval. Therefore the inequalities (\ref{ineq9})-(\ref{ineq12}) trivially hold
\be
0=0\ , \quad
0=0\ , \quad
{c\over \alpha^2}\geq 0\ ,\quad
c\geq0\ .
\ee

{\it $d=4, {\cal N}=1$}\, SCFT~~~For these theories, there is a conserved $U(1)$ R-symmetry. We consider Lagrangian theories in flat space with the entangling surface being a round 2-sphere with radius $R$, $\bar S_\alpha$ enjoys a universal behavior at $\alpha\ll 1$~\cite{Zhou:2015cpa}
\be\label{n1s}
\bar S_{\alpha\ll1} = {4\over 27\alpha^2}(3c-2a){V_3\over 2\pi}\ ,\quad V_3=-2\pi\log{R\over \epsilon}\ ,
\ee where $V_3$ is the regularized volume of $\Bbb{H}^3$ and $a,c$ are the Weyl anomaly coefficients defined from the anomalous trace of the stress tensor in $4d$ curved background,
\be
\langle T_\mu^{\,\mu}\rangle = {1\over (4\pi)^2}(a E - c W^2)\ .
\ee
The formula (\ref{n1s}) was derived from the free field computation with a nontrivial R-symmetry background and shown~\cite{Zhou:2015cpa} to be universal for SCFTs by matching to the $4d$ supersymmetric Casimir energy~\cite{Assel:2015nca} on an extremely squashed sphere. Plugging (\ref{n1s}) into the four different inequalities (\ref{ineq9})-(\ref{ineq12}), one obtains a single constraint $
3c-2a\geq 0$, which is the Hofman-Maldacena upper bound for general ${\cal N}=1$ SCFTs~\cite{Hofman:2008ar}. Together with the unitarity bound $c>0$ and the positivity of the universal spherical entanglement entropy $S_{\text{EE}}\propto a$~\cite{Casini:2011kv,positivea}, we have
\be
{3\over 2}\geq{a\over c}\geq 0\ .
\ee Notice that this is not as tight as the ${\cal N}=1$ Hofman-Maldacena bounds, ${1\over 2}\leq {a\over c}\leq {3\over 2}$. For recent approaches to a proof of Hofman-Maldacena bounds, see~\cite{Hofman:2016awc,Komargodski:2016gci}. 

{\it $d=4, {\cal N}=2$}\, SCFT~~~For these theories, the R-symmetry is $SU(2)_R\times U(1)_R$. The $U(1)_R$ may be broken for the purpose of defining sphere partition functions~\cite{Gerchkovitz:2014gta}. We turn on the background field corresponding to $U(1)_J\subset SU(2)_R$ to twist the R\'enyi entropy. Notice that we focus on the universal logarithmic term of the supersymmetric R\'enyi entropy. For Lagrangian theories, $\bar S_\alpha$ has been determined completely in terms of $4d$ Weyl anomaly coefficients $a,c$~\cite{Zhou:2015cpa}
\be\label{n2s}
\bar S_\alpha =\left({c\over \alpha}+4a-c\right){V_3\over 2\pi}\ .
\ee This result was first derived from the free field computation and shown to be universal~\cite{Zhou:2015cpa} by matching to the localization results in~\cite{HZ,Hama:2012bg}. Plugging (\ref{n2s}) into the inequalities (\ref{ineq9})-(\ref{ineq12}) one obtains
\be
c\geq 0\ ,\quad
c\geq 0\ ,\quad
c+(2a-c)\alpha\geq 0\ ,\quad
4a-c+{c\over \alpha^2}\geq 0\ .
\ee
The large $\alpha$ limit of the third inequality gives $2a-c\geq 0$, which is the Hofman-Maldacena lower bound for general ${\cal N}=2$ SCFTs. Together with $a/c\leq 3/2$ one obtains
\be
{1\over 2}\leq {a\over c}\leq {3\over 2}\label{n2low}\ .
\ee The upper bound comes from describing ${\cal N}=2$ theories as ${\cal N}=1$ ones. Notice that (\ref{n2low}) is not as tight as the ${\cal N}=2$ Hofman-Maldacena bounds, ${1\over 2}\leq {a\over c}\leq {5\over 4}$. Four-dimensional ${\cal N}=4$ Super-Yang Mills (SYM) always has positive $a=c$ and its universal supersymmetric R\'enyi entropy can be derived either from the free field computation or from the holographic computation on $5d$ BPS charged topological AdS black holes~\cite{HZ}.

{\it $d=6, {\cal N}=(2,0)$}\,SCFT~~~For these theories, the R-symmetry group is $SO(5)$ and the two Cartans are on the equal footing. $\bar S_\alpha$ has been determined completely~\cite{Zhou:2015kaj} in terms of the $6d$ Weyl anomaly coefficients $a,c$~\cite{Beem:2014kka,Cordova:2015vwa}, which are defined from the anomalous trace of the $6d$ stress tensor in curved background (with the normalization such that a free tensor multiplet has unit $a$ and $c$)
\ba
\bar S_\alpha {\pi^2\over V_5}&= {r_1^2r_2^2\over 12}{7a-3c\over 4}\left(\gamma-1\right)^3 \nonumber\\
&+ {r_1r_2\over 12}c\left(\gamma-1\right)^2 +{1+2r_1r_2\over 12}c\left(\gamma-1\right) +{7\over 12}a\ ,\label{6dsre}\ea where $\gamma:=1/\alpha$ and $r_{1,2}\geq 0$ are the weights of the two chemical potentials with a constraint $r_1+r_2=1$. This result was obtained by making use of $\bar S_\alpha$ of a free tensor multiplet~\cite{Nian:2015xky}, the 2- and 3-point functions of the stress tensor multiplet~\cite{Beem:2014kka} and the $6d$ supersymmetric Casimir energy~\cite{Bobev:2015kza} on an extremely squashed sphere. The large $N$ limit of (\ref{6dsre}) agrees with the holographic result from $7d$ BPS charged topological AdS black holes~\cite{Zhou:2015kaj}. Plugging (\ref{6dsre}) into (\ref{ineq9})-(\ref{ineq12}) and demanding that the inequalities hold for any positive $\alpha$, one can get
\be
{a\over c}\geq {3\over 7}\ ,\quad c\geq 0\ .
\ee The lower bound of $a/c$ together with the unitarity bound $c>0$ also proves the positivity of $a$. 

\subsection{Other examples}
In odd dimensional CFTs, the finite parts of the entanglement entropy and the R\'enyi entropy (or supersymmetric) associated with a spherical entangling surface in flat space are considered to be universal and physical. One can compute them by mapping to a branched sphere $\Bbb{S}^d_\alpha$ because there is no conformal anomaly. For $d=3, {\cal N}=2$ superconformal Chern-Simons gauge theories with M-theory duals, $\bar S_\alpha$ in the large $N$ limit has the scaling $\bar S_\alpha/\bar S_1=(3\alpha+1)/4\alpha$, which satisfies all the four inequalities (\ref{ineq9})-(\ref{ineq12}) as observed in~\cite{NY}. For $d=5,{\cal N}=1$ superconformal theory with AdS$_6$ dual~\cite{Brandhuber:1999np}, $\bar S_\alpha$ in the large $N$ limit has the scaling~\cite{5dsre,Alday:2014fsa,Hama:2014iea} $\bar S_\alpha/\bar S_1=(19\alpha^2+7\alpha+1)/27\alpha^2$, which also satisfies the four inequalities as observed in~\cite{Hama:2014iea}. One can also numerically check the inequalities for other $5d$ or $3d$ superconformal examples including ABJM with finite $N$.

%%%%%%%%%%%%%%%%%%%%%%%%%%%%%%%%%%%%%%%%%%%%%%%%%%
\se{Acknowledgements}

\begin{acknowledgments}
I would like to thank Zohar Komargodski, Amit Sever and Jacob Sonnenschein for insightful discussions and Zohar Komargodski for reading the manuscript. This work was supported by ``The PBC program of the Israel council of higher education'' and in part by the Israel Science Foundation (grant 1989/14), the US-Israel bi-national fund (BSF) grant 2012383 and the German Israel bi-national fund GIF grant number I-244-303.7-2013.
\end{acknowledgments}
%%%%%%%%%%%%%%%%%%%%%%%%%%%%%%%%%%%%%%%%%%%%%%%%%%
\appendix
\se{A holographic derivation of $S\geq 2\pi Q$}
We consider a $d+1$-dimensional BPS charged topological AdS black hole, which is the gravity dual of the ground state in SCFT$_d$ on supersymmetric $\Bbb{S}_\alpha^1\times\Bbb{H}^{d-1}$ and used to compute the holographic supersymmetric R\'enyi entropy. Below we will take $5d$ ${\cal N}=1$ supersymmetric $USp(2N)$ gauge theory with $N_f$ fundamental hypermultiplets and a single hypermultiplet in the antisymmetric representation as an example, but the argument also goes well in other dimensions. The gravity dual of the ground state of this $5d$ SCFT on $\Bbb{S}^1_\alpha\times\Bbb{H}^4$ is given by a $6d$ BPS charged topological AdS black hole~\cite{5dsre,Alday:2014fsa,Hama:2014iea,Cvetic:1999un}
\ba
ds^2 &= -H^{-3/2}f dt^2 + H^{1/2}(f^{-1}dr^2 + r^2 d\Omega_{4,-1}^2) \nn\\  f &= -1 + {r^2\over R^2}H^2\ ,\quad H= 1 + {q\over r^3}\ ,\label{6dmetric}
\ea together with the scalar and the gauge field
\be
X= H^{-1/4}\ ,~A = \left(\sqrt{2}(H^{-1} - 1) + \mu\right)d\tau\ ,
\ee where $d\Omega_{4,-1}^2$ denotes the metric on $\Bbb{H}^4$ and $t=-i\tau$. We define a rescaled charge $\kappa=q/r_h^3$, where the event horizon $r_h$ is the largest root of the equation $f(r_h)=0$. The Hawking temperature, the Bekenstein-Hawking entropy, the total charge and the chemical potential can be worked out as follows,
\ba
&T={1 \over 2\pi R}{2-\kappa\over2(1+\kappa)^2}\ ,\\
&S={V_4 R \over 4 G_6}r_h^3\ ,\\
&Q=-3\sqrt{2}\kappa{V_4\,r_h^3\over 16\pi G_6}\ ,\\
&\mu={\sqrt{2}\over \kappa^{-1}+1}\ ,
\ea where $G_6$ is the $6$-dimensional Newton constant and $V_4$ is the regularized volume of unit $\Bbb{H}^4$.
We choose a new normalization for $\mu$ and $Q$ such that when $T={1\over 2\pi R\alpha}$, $\widetilde\mu$ takes the value ${\alpha-1\over \alpha}$ matching to that in (\ref{muv}). In this case, the normalized charge is given by
\be
\widetilde Q = -\kappa{V_4\,r_h^3\over 8\pi G_6}\ .
\ee
The horizon radius $r_h$ should be positive, $r_h> 0$. Then the positivity of
\be
S-2\pi R \widetilde Q = {V_4 R \over 4 G_6}r_h^3 H(r_h)
\ee is guaranteed by the causality, since the sign flip of $H(r)$ in the metric (\ref{6dmetric}) is forbidden before reaching to the horizon. $H(r_h)\geq 0$ ensures $S\geq 2\pi R \widetilde Q$. The same argument goes well in other dimensions, $d=3,4,6$. Notice that we restored the length scale $R$, which has been omitted in the body part. The holographic supersymmetric R\'enyi entropy can be computed straightforwardly by employing the formula derived in~\cite{Huang:2014gca}. Recently there is a holographic study of the R\'enyi entropy inequalities~\cite{Nakaguchi:2016zqi} based on~\cite{Dong:2016fnf}, it would be interesting to consider our bound $S\geq 2\pi R Q$ along that way. 

%%%%%%%%%%%%%%%%%%%%%%%%%%%%%%%%%%%%%%%%%%%%%%%%%%
%\bibliographystyle{apsrev4-1}


\begin{thebibliography}{100}

%\cite{Vidal:2002rm}
\bibitem{Vidal:2002rm} 
  G.~Vidal, J.~I.~Latorre, E.~Rico and A.~Kitaev,
  %``Entanglement in quantum critical phenomena,''
  Phys.\ Rev.\ Lett.\  {\bf 90}, 227902 (2003)
%  doi:10.1103/PhysRevLett.90.227902
%  [quant-ph/0211074].
  %%CITATION = doi:10.1103/PhysRevLett.90.227902;%%
  %306 citations counted in INSPIRE as of 29 Jun 2016
  
 %\cite{Kitaev:2005dm}
\bibitem{Kitaev:2005dm} 
  A.~Kitaev and J.~Preskill,
  %``Topological entanglement entropy,''
  Phys.\ Rev.\ Lett.\  {\bf 96}, 110404 (2006)
%  doi:10.1103/PhysRevLett.96.110404
%  [hep-th/0510092].
  %%CITATION = doi:10.1103/PhysRevLett.96.110404;%%
  %398 citations counted in INSPIRE as of 29 Jun 2016
  
   %\cite{Kitaev:2005dm}
\bibitem{Levin:2006} 
  M.~Levin and X.G.~Wen,
  %``Topological entanglement entropy,''
  Phys.\ Rev.\ Lett.\  {\bf 96}, 110405 (2006)
%  doi:10.1103/PhysRevLett.96.110404
%  [hep-th/0510092].
  %%CITATION = doi:10.1103/PhysRevLett.96.110404;%%
  %398 citations counted in INSPIRE as of 29 Jun 2016
  
 %\cite{Casini:2004bw}
\bibitem{Casini:2004bw} 
  H.~Casini and M.~Huerta,
  %``A Finite entanglement entropy and the c-theorem,''
  Phys.\ Lett.\ B {\bf 600}, 142 (2004)
%  doi:10.1016/j.physletb.2004.08.072
%  [hep-th/0405111].
  %%CITATION = doi:10.1016/j.physletb.2004.08.072;%%
  %142 citations counted in INSPIRE as of 29 Jun 2016

  %\cite{Casini:2006es}
\bibitem{Casini:2006es} 
  H.~Casini and M.~Huerta,
  %``A c-theorem for the entanglement entropy,''
  J.\ Phys.\ A {\bf 40}, 7031 (2007)
%  doi:10.1088/1751-8113/40/25/S57
%  [cond-mat/0610375].
  %%CITATION = doi:10.1088/1751-8113/40/25/S57;%%
  %48 citations counted in INSPIRE as of 30 Jun 2016
  
  %\cite{Casini:2012ei}
\bibitem{Casini:2012ei} 
  H.~Casini and M.~Huerta,
  %``On the RG running of the entanglement entropy of a circle,''
  Phys.\ Rev.\ D {\bf 85}, 125016 (2012)
%  doi:10.1103/PhysRevD.85.125016
%  [arXiv:1202.5650 [hep-th]].
  %%CITATION = doi:10.1103/PhysRevD.85.125016;%%
  %143 citations counted in INSPIRE as of 29 Jun 2016
  
%\cite{Casini:2008cr}
\bibitem{Casini:2008cr} 
  H.~Casini,
  %``Relative entropy and the Bekenstein bound,''
  Class.\ Quant.\ Grav.\  {\bf 25}, 205021 (2008)
%  doi:10.1088/0264-9381/25/20/205021
%  [arXiv:0804.2182 [hep-th]].
  %%CITATION = doi:10.1088/0264-9381/25/20/205021;%%
  %42 citations counted in INSPIRE as of 14 Jul 2016
  
 %\cite{Ryu:2006bv}
\bibitem{Ryu:2006bv} 
  S.~Ryu and T.~Takayanagi,
  %``Holographic derivation of entanglement entropy from AdS/CFT,''
  Phys.\ Rev.\ Lett.\  {\bf 96}, 181602 (2006)
%  doi:10.1103/PhysRevLett.96.181602
%  [hep-th/0603001].
  %%CITATION = doi:10.1103/PhysRevLett.96.181602;%%
  %994 citations counted in INSPIRE as of 29 Jun 2016
  
%\cite{Callan:1994py}
\bibitem{Callan:1994py} 
  C.~G.~Callan, Jr. and F.~Wilczek,
  %``On geometric entropy,''
  Phys.\ Lett.\ B {\bf 333}, 55 (1994)
%  doi:10.1016/0370-2693(94)91007-3
%  [hep-th/9401072].
  %%CITATION = doi:10.1016/0370-2693(94)91007-3;%%
  %416 citations counted in INSPIRE as of 16 Jul 2016
  
 %\cite{Solodukhin:2008dh}
\bibitem{Solodukhin:2008dh} 
  S.~N.~Solodukhin,
  %``Entanglement entropy, conformal invariance and extrinsic geometry,''
  Phys.\ Lett.\ B {\bf 665}, 305 (2008)
%  doi:10.1016/j.physletb.2008.05.071
%  [arXiv:0802.3117 [hep-th]].
  %%CITATION = doi:10.1016/j.physletb.2008.05.071;%%
  %152 citations counted in INSPIRE as of 29 Jun 2016
  
 %\cite{Myers:2010tj}
\bibitem{Myers:2010tj} 
  R.~C.~Myers and A.~Sinha,
  %``Holographic c-theorems in arbitrary dimensions,''
  JHEP {\bf 1101}, 125 (2011)
%  doi:10.1007/JHEP01(2011)125
%  [arXiv:1011.5819 [hep-th]].
  %%CITATION = doi:10.1007/JHEP01(2011)125;%%
  %260 citations counted in INSPIRE as of 29 Jun 2016
  
  %\cite{Casini:2011kv}
\bibitem{Casini:2011kv} 
  H.~Casini, M.~Huerta and R.~C.~Myers,
  %``Towards a derivation of holographic entanglement entropy,''
  JHEP {\bf 1105}, 036 (2011)
%  doi:10.1007/JHEP05(2011)036
%  [arXiv:1102.0440 [hep-th]].
  %%CITATION = doi:10.1007/JHEP05(2011)036;%%
  %374 citations counted in INSPIRE as of 29 Jun 2016

%\cite{Jafferis:2011zi}
\bibitem{Jafferis:2011zi} 
  D.~L.~Jafferis, I.~R.~Klebanov, S.~S.~Pufu and B.~R.~Safdi,
  %``Towards the F-Theorem: N=2 Field Theories on the Three-Sphere,''
  JHEP {\bf 1106}, 102 (2011)
%  doi:10.1007/JHEP06(2011)102
%  [arXiv:1103.1181 [hep-th]].
  %%CITATION = doi:10.1007/JHEP06(2011)102;%%
  %221 citations counted in INSPIRE as of 29 Jun 2016  
  
 %\cite{Liu:2012eea}
\bibitem{Liu:2012eea} 
  H.~Liu and M.~Mezei,
  %``A Refinement of entanglement entropy and the number of degrees of freedom,''
  JHEP {\bf 1304}, 162 (2013)
%  doi:10.1007/JHEP04(2013)162
%  [arXiv:1202.2070 [hep-th]].
  %%CITATION = doi:10.1007/JHEP04(2013)162;%%
  %104 citations counted in INSPIRE as of 29 Jun 2012
  
 %\cite{Faulkner:2016mzt}
\bibitem{Faulkner:2016mzt} 
  T.~Faulkner, R.~G.~Leigh, O.~Parrikar and H.~Wang,
  %``Modular Hamiltonians for Deformed Half-Spaces and the Averaged Null Energy Condition,''
  arXiv:1605.08072 [hep-th].
  %%CITATION = ARXIV:1605.08072;%%
  %2 citations counted in INSPIRE as of 24 Aug 2016
  
 %\cite{Lashkari:2016idm}
\bibitem{Lashkari:2016idm} 
  N.~Lashkari, J.~Lin, H.~Ooguri, B.~Stoica and M.~Van Raamsdonk,
  %``Gravitational Positive Energy Theorems from Information Inequalities,''
  arXiv:1605.01075 [hep-th].
  %%CITATION = ARXIV:1605.01075;%%
  %2 citations counted in INSPIRE as of 16 Jul 2016

\bibitem{BeckSchlogl}
Christian Beck; Friedrich Schl\"{o}gl, 
{\it Thermodynamics of chaotic systems: an introduction,}
Cambridge Univ. Press (1995).

\bibitem{quantumP}
To generalize the proof of classical information theoretic inequalities to quantum mechanical ones, one simply diagonalizes density matrices $\rho, \sigma$ with unitary matrices, which does not change the R\'enyi entropy (or R\'enyi divergence). 

\bibitem{HR}
  L.~Y.~Hung, R.~C.~Myers, M.~Smolkin and A.~Yale,
%  ``Holographic Calculations of Renyi Entropy,''
  JHEP {\bf 1112}, 047 (2011)
%  doi:10.1007/JHEP12(2011)047
%  [arXiv:1110.1084 [hep-th]].
  %%CITATION = doi:10.1007/JHEP12(2011)047;%%
  %79 citations counted in INSPIRE as of 14 Jun 2016

%\cite{Calabrese:2004eu}
\bibitem{Calabrese:2004eu} 
  P.~Calabrese and J.~L.~Cardy,
  %``Entanglement entropy and quantum field theory,''
  J.\ Stat.\ Mech.\  {\bf 0406}, P06002 (2004)
%  doi:10.1088/1742-5468/2004/06/P06002
%  [hep-th/0405152].
  %%CITATION = doi:10.1088/1742-5468/2004/06/P06002;%%
  %502 citations counted in INSPIRE as of 16 Jul 2016
  
 %\cite{Calabrese:2009qy}
\bibitem{Calabrese:2009qy} 
  P.~Calabrese and J.~Cardy,
  %``Entanglement entropy and conformal field theory,''
  J.\ Phys.\ A {\bf 42}, 504005 (2009)
%  doi:10.1088/1751-8113/42/50/504005
%  [arXiv:0905.4013 [cond-mat.stat-mech]].
  %%CITATION = doi:10.1088/1751-8113/42/50/504005;%%
  %253 citations counted in INSPIRE as of 16 Jul 2016

%\cite{Casini:2010kt}
\bibitem{Casini:2010kt} 
  H.~Casini and M.~Huerta,
  %``Entanglement entropy for the n-sphere,''
  Phys.\ Lett.\ B {\bf 694}, 167 (2011)
%  doi:10.1016/j.physletb.2010.09.054
%  [arXiv:1007.1813 [hep-th]].
  %%CITATION = doi:10.1016/j.physletb.2010.09.054;%%
  %62 citations counted in INSPIRE as of 16 Jul 2016
  
%\cite{Klebanov:2011uf}
\bibitem{Klebanov:2011uf} 
  I.~R.~Klebanov, S.~S.~Pufu, S.~Sachdev and B.~R.~Safdi,
  %``Renyi Entropies for Free Field Theories,''
  JHEP {\bf 1204}, 074 (2012)
%  doi:10.1007/JHEP04(2012)074
%  [arXiv:1111.6290 [hep-th]].
  %%CITATION = doi:10.1007/JHEP04(2012)074;%%
  %53 citations counted in INSPIRE as of 16 Jul 2016
  
 %\cite{Fursaev:2012mp}
\bibitem{Fursaev:2012mp} 
  D.~V.~Fursaev,
  %``Entanglement Renyi Entropies in Conformal Field Theories and Holography,''
  JHEP {\bf 1205}, 080 (2012)
%  doi:10.1007/JHEP05(2012)080
%  [arXiv:1201.1702 [hep-th]].
  %%CITATION = doi:10.1007/JHEP05(2012)080;%%
  %36 citations counted in INSPIRE as of 16 Jul 2016
  
  %\cite{Dowker:2012rp}
\bibitem{Dowker:2012rp} 
  J.~S.~Dowker,
  %``Sphere Renyi entropies,''
  J.\ Phys.\ A {\bf 46}, 225401 (2013)
%  doi:10.1088/1751-8113/46/22/225401
%  [arXiv:1212.2098 [hep-th]].
  %%CITATION = doi:10.1088/1751-8113/46/22/225401;%%
  %7 citations counted in INSPIRE as of 16 Jul 2016
  
 %\cite{Lee:2014zaa}
\bibitem{Lee:2014zaa} 
  J.~Lee, A.~Lewkowycz, E.~Perlmutter and B.~R.~Safdi,
  %``Rényi entropy, stationarity, and entanglement of the conformal scalar,''
  JHEP {\bf 1503}, 075 (2015)
%  doi:10.1007/JHEP03(2015)075
%  [arXiv:1407.7816 [hep-th]].
  %%CITATION = doi:10.1007/JHEP03(2015)075;%%
  %24 citations counted in INSPIRE as of 16 Jul 2016

\bibitem{NY}
  T.Nishioka and I.Yaakov, 
%  ``Supersymmetric Renyi Entropy,'' 
  JHEP {\bf 1310}, 155 (2013) 
%  doi:10.1007/JHEP10(2013)155 
%  [arXiv:1306.2958 [hep-th]].
  %%CITATION = doi:10.1007/JHEP10(2013)155;%%
  %25 citations counted in INSPIRE as of 01 Dec 2015
  
   %\cite{Huang:2014gca}
\bibitem{Huang:2014gca} 
  X.~Huang, S.~J.~Rey and Y.~Zhou,
  %``Three-dimensional SCFT on conic space as hologram of charged topological black hole,''
  JHEP {\bf 1403}, 127 (2014)
%  doi:10.1007/JHEP03(2014)127
%  [arXiv:1401.5421 [hep-th]].
  %%CITATION = doi:10.1007/JHEP03(2014)127;%%
  %18 citations counted in INSPIRE as of 06 Jul 2016
  
\bibitem{Renyi}
A.R\'enyi,
Proceedings of the Fourth Berkeley Symposium on Mathematical Statistics and Probability, vol. 1, pp. 547-561, 1961.

 \bibitem{RDivergence}
T. van Erven and P. Harremo\"{e}s,
%Renyi Divergence and Kullback-Leibler Divergence, 
{\it IEEE Trans. Inf. Theory,} vol. 60, no. 7, pp. 3797-3820, 2014
  
 \bibitem{HZ}
  X.~Huang and Y.~Zhou,
%  ``$ \mathcal{N}=4 $ Super-Yang-Mills on conic space as hologram of STU topological black hole,''
  JHEP {\bf 1502}, 068 (2015)
%  doi:10.1007/JHEP02(2015)068
%  [arXiv:1408.3393 [hep-th]].
  
 %cite{Crossley:2014oea}
\bibitem{Crossley:2014oea} 
  M.~Crossley, E.~Dyer and J.~Sonner,
%  ``Super-Renyi entropy, Wilson loops for N=4 SYM and their gravity duals,''
  JHEP {\bf 1412}, 001 (2014)
%  doi:10.1007/JHEP12(2014)001
%  [arXiv:1409.0542 [hep-th]].
  %%CITATION = doi:10.1007/JHEP12(2014)001;%%
  %10 citations counted in INSPIRE as of 01 Dec 2015

\bibitem{5dsre} 
  S.~J.~Rey and Y.~Zhou,
  unpublished notes
  
%\cite{Alday:2014fsa}
\bibitem{Alday:2014fsa} 
  L.~F.~Alday, P.~Richmond and J.~Sparks,
%  ``The holographic supersymmetric Renyi entropy in five dimensions,''
  JHEP {\bf 1502}, 102 (2015)
%  doi:10.1007/JHEP02(2015)102
%  [arXiv:1410.0899 [hep-th]].
  %%CITATION = doi:10.1007/JHEP02(2015)102;%%
  %7 citations counted in INSPIRE as of 01 Dec 2015
  
  %\cite{Hama:2014iea}
\bibitem{Hama:2014iea} 
  N.~Hama, T.~Nishioka and T.~Ugajin,
%  ``Supersymmetric Renyi entropy in five dimensions,''
  JHEP {\bf 1412}, 048 (2014)
%  doi:10.1007/JHEP12(2014)048
%  [arXiv:1410.2206 [hep-th]].
  %%CITATION = doi:10.1007/JHEP12(2014)048;%%
  %6 citations counted in INSPIRE as of 01 Dec 2015
  
  \bibitem{Zhou:2015cpa} 
  Y.~Zhou,
%  ``Universal Features of Four-Dimensional Superconformal Field Theory on Conic Space,''
  JHEP {\bf 1508}, 052 (2015)
%  doi:10.1007/JHEP08(2015)052
%  [arXiv:1506.06512 [hep-th]].
  %%CITATION = doi:10.1007/JHEP08(2015)052;%%
  %2 citations counted in INSPIRE as of 01 Dec 2015
  
 %\cite{Nian:2015xky}
\bibitem{Nian:2015xky} 
  J.~Nian and Y.~Zhou,
  %``Rényi entropy of a free (2, 0) tensor multiplet and its supersymmetric counterpart,''
  Phys.\ Rev.\ D {\bf 93}, no. 12, 125010 (2016)
%  doi:10.1103/PhysRevD.93.125010
%  [arXiv:1511.00313 [hep-th]].
  %%CITATION = doi:10.1103/PhysRevD.93.125010;%%
  %1 citations counted in INSPIRE as of 07 Jul 2016
  
 %\cite{Zhou:2015kaj}
\bibitem{Zhou:2015kaj} 
  Y.~Zhou,
%  ``Supersymmetric Renyi Entropy and Weyl Anomalies in Six-Dimensional (2,0) Theories,''
  JHEP {\bf 1606}, 064 (2016)
%  doi:10.1007/JHEP06(2016)064
%  [arXiv:1512.03008 [hep-th]].
  %%CITATION = doi:10.1007/JHEP06(2016)064;%%
  
 %\cite{Mori:2015bro}
\bibitem{Mori:2015bro} 
  H.~Mori,
%  ``Supersymmetric Renyi entropy in two dimensions,''
  JHEP {\bf 1603}, 058 (2016)
%  doi:10.1007/JHEP03(2016)058
%  [arXiv:1512.02829 [hep-th]].
  %%CITATION = doi:10.1007/JHEP03(2016)058;%%
  
 %\cite{Giveon:2015cgs}
\bibitem{Giveon:2015cgs} 
  A.~Giveon and D.~Kutasov,
  %``Supersymmetric Renyi entropy in CFT$_{2}$ and AdS$_{3}$,''
  JHEP {\bf 1601}, 042 (2016)
%  doi:10.1007/JHEP01(2016)042
%  [arXiv:1510.08872 [hep-th]].
  %%CITATION = doi:10.1007/JHEP01(2016)042;%%
  %4 citations counted in INSPIRE as of 07 Jul 2016

\bibitem{testtrace}
This trace formula for supersymmetric R\'enyi entropy has passed nontrivial tests. For instance one can check a relation derived from it~\cite{Zhou:2015kaj},
$$S^\prime_{\alpha=1}=-V_{d-1}\left({\pi^{{d\over 2}+1}\Gamma({d\over 2})(d-1)\over (d+1)!}C_T -{\pi^{d+3\over 2}\over 2^{d-3}(d-1)\Gamma({d-1\over 2})}C_J\right),$$ where $C_T$ and $C_J$ are defined from the stress tensor 2-point correlator and the R-current 2-point correlator, respectively. Notice that $\hat Q$ in this letter is equal to $\alpha \hat Q$ in~\cite{Zhou:2015kaj}. The charged R\'enyi entropy defined in~\cite{Belin:2013uta} is not supersymmetric.

%\cite{Belin:2013uta}
\bibitem{Belin:2013uta} 
  A.~Belin, L.~Y.~Hung, A.~Maloney, S.~Matsuura, R.~C.~Myers and T.~Sierens,
  %``Holographic Charged Renyi Entropies,''
  JHEP {\bf 1312}, 059 (2013)
%  doi:10.1007/JHEP12(2013)059
%  [arXiv:1310.4180 [hep-th]].
  %%CITATION = doi:10.1007/JHEP12(2013)059;%%
  %29 citations counted in INSPIRE as of 16 Jul 2016
  
\bibitem{renyiRelative}
Both $K$ and $\hat Q^\prime$ are hermitian in real time quantization. The chemical potential is kept to be real in our convention and it is unchanged under Weyl transformations. 
%$\hat Q$ is anti-hermitian following (A.21) in~\cite{Zhou:2015kaj}. 
A path integral approach to R\'enyi divergence of different distributions is given in~\cite{Lashkari:2014yva,Sarosi:2016oks}.

%\cite{Lashkari:2014yva}
\bibitem{Lashkari:2014yva} 
  N.~Lashkari,
  %``Relative Entropies in Conformal Field Theory,''
  Phys.\ Rev.\ Lett.\  {\bf 113}, 051602 (2014)
%  doi:10.1103/PhysRevLett.113.051602
%  [arXiv:1404.3216 [hep-th]].
  %%CITATION = doi:10.1103/PhysRevLett.113.051602;%%
  %6 citations counted in INSPIRE as of 16 Jul 2016
  
%\cite{Sarosi:2016oks}
\bibitem{Sarosi:2016oks} 
  G.~S\'arosi and T.~Ugajin,
  %``Relative Entropy of Excited States in Two Dimensional Conformal Field Theories,''
  arXiv:1603.03057 [hep-th].
  %%CITATION = ARXIV:1603.03057;%%
  %2 citations counted in INSPIRE as of 16 Jul 2016
  
 %\cite{Assel:2015nca}
\bibitem{Assel:2015nca} 
  B.~Assel, D.~Cassani, L.~Di Pietro, Z.~Komargodski, J.~Lorenzen and D.~Martelli,
  %``The Casimir Energy in Curved Space and its Supersymmetric Counterpart,''
  JHEP {\bf 1507}, 043 (2015)
%  doi:10.1007/JHEP07(2015)043
%  [arXiv:1503.05537 [hep-th]].
  %%CITATION = doi:10.1007/JHEP07(2015)043;%%
  %28 citations counted in INSPIRE as of 07 Jul 2016
  
 %\cite{Hofman:2008ar}
\bibitem{Hofman:2008ar} 
  D.~M.~Hofman and J.~Maldacena,
  %``Conformal collider physics: Energy and charge correlations,''
  JHEP {\bf 0805}, 012 (2008)
%  doi:10.1088/1126-6708/2008/05/012
%  [arXiv:0803.1467 [hep-th]].
  %%CITATION = doi:10.1088/1126-6708/2008/05/012;%%
  %267 citations counted in INSPIRE as of 16 Jul 2016
  
 \bibitem{positivea}
 For the entangling surface with a nontrivial topology, this positivity is not guaranteed~\cite{Perlmutter:2015vma}.
 
 %\cite{Perlmutter:2015vma}
\bibitem{Perlmutter:2015vma} 
  E.~Perlmutter, M.~Rangamani and M.~Rota,
  %``Central Charges and the Sign of Entanglement in 4D Conformal Field Theories,''
  Phys.\ Rev.\ Lett.\  {\bf 115}, no. 17, 171601 (2015)
%  doi:10.1103/PhysRevLett.115.171601
%  [arXiv:1506.01679 [hep-th]].
  %%CITATION = doi:10.1103/PhysRevLett.115.171601;%%
  %16 citations counted in INSPIRE as of 17 Jul 2016
  
  %\cite{Hofman:2016awc}
\bibitem{Hofman:2016awc} 
  D.~M.~Hofman, D.~Li, D.~Meltzer, D.~Poland and F.~Rejon-Barrera,
  %``A Proof of the Conformal Collider Bounds,''
  JHEP {\bf 1606}, 111 (2016)
%  doi:10.1007/JHEP06(2016)111
%  [arXiv:1603.03771 [hep-th]].
  %%CITATION = doi:10.1007/JHEP06(2016)111;%%
  %9 citations counted in INSPIRE as of 16 Jul 2016
  
%\cite{Komargodski:2016gci}
\bibitem{Komargodski:2016gci} 
  Z.~Komargodski, M.~Kulaxizi, A.~Parnachev and A.~Zhiboedov,
  %``Conformal Field Theories and Deep Inelastic Scattering,''
  arXiv:1601.05453 [hep-th].
  %%CITATION = ARXIV:1601.05453;%%
  %7 citations counted in INSPIRE as of 16 Jul 2016
  
%\cite{Gerchkovitz:2014gta}
\bibitem{Gerchkovitz:2014gta} 
  E.~Gerchkovitz, J.~Gomis and Z.~Komargodski,
  %``Sphere Partition Functions and the Zamolodchikov Metric,''
  JHEP {\bf 1411}, 001 (2014)
%  doi:10.1007/JHEP11(2014)001
%  [arXiv:1405.7271 [hep-th]].
  %%CITATION = doi:10.1007/JHEP11(2014)001;%%
  %38 citations counted in INSPIRE as of 07 Jul 2016
  
%\cite{Hama:2012bg}
\bibitem{Hama:2012bg} 
  N.~Hama and K.~Hosomichi,
  %``Seiberg-Witten Theories on Ellipsoids,''
  JHEP {\bf 1209}, 033 (2012)
  Addendum: [JHEP {\bf 1210}, 051 (2012)]
%  doi:10.1007/JHEP09(2012)033, 10.1007/JHEP10(2012)051
%  [arXiv:1206.6359 [hep-th]].
  %%CITATION = doi:10.1007/JHEP09(2012)033, 10.1007/JHEP10(2012)051;%%
  %85 citations counted in INSPIRE as of 07 Jul 2016
  
%\cite{Beem:2014kka}
\bibitem{Beem:2014kka} 
  C.~Beem, L.~Rastelli and B.~C.~van Rees,
  %``$ \mathcal{W} $ symmetry in six dimensions,''
  JHEP {\bf 1505}, 017 (2015)
%  doi:10.1007/JHEP05(2015)017
%  [arXiv:1404.1079 [hep-th]].
  %%CITATION = doi:10.1007/JHEP05(2015)017;%%
  %35 citations counted in INSPIRE as of 07 Jul 2016
  
%\cite{Cordova:2015vwa}
\bibitem{Cordova:2015vwa} 
  C.~Cordova, T.~T.~Dumitrescu and X.~Yin,
  %``Higher Derivative Terms, Toroidal Compactification, and Weyl Anomalies in Six-Dimensional (2,0) Theories,''
  arXiv:1505.03850 [hep-th].
  %%CITATION = ARXIV:1505.03850;%%
  %19 citations counted in INSPIRE as of 07 Jul 2016
  
%\cite{Bobev:2015kza}
\bibitem{Bobev:2015kza} 
  N.~Bobev, M.~Bullimore and H.~C.~Kim,
  %``Supersymmetric Casimir Energy and the Anomaly Polynomial,''
  JHEP {\bf 1509}, 142 (2015)
%  doi:10.1007/JHEP09(2015)142
%  [arXiv:1507.08553 [hep-th]].
  %%CITATION = doi:10.1007/JHEP09(2015)142;%%
  %8 citations counted in INSPIRE as of 07 Jul 2016
  
%\cite{Brandhuber:1999np}
\bibitem{Brandhuber:1999np} 
  A.~Brandhuber and Y.~Oz,
  %``The D-4 - D-8 brane system and five-dimensional fixed points,''
  Phys.\ Lett.\ B {\bf 460}, 307 (1999)
%  doi:10.1016/S0370-2693(99)00763-7
%  [hep-th/9905148].
  %%CITATION = doi:10.1016/S0370-2693(99)00763-7;%%
  %86 citations counted in INSPIRE as of 08 Jul 2016
  
 %\cite{Cvetic:1999un}
\bibitem{Cvetic:1999un} 
  M.~Cvetic, H.~Lu and C.~N.~Pope,
  %``Gauged six-dimensional supergravity from massive type IIA,''
  Phys.\ Rev.\ Lett.\  {\bf 83}, 5226 (1999)
%  doi:10.1103/PhysRevLett.83.5226
%  [hep-th/9906221].
  %%CITATION = doi:10.1103/PhysRevLett.83.5226;%%
  %94 citations counted in INSPIRE as of 18 Jul 2016
  
 %\cite{Dong:2016fnf}
\bibitem{Dong:2016fnf} 
  X.~Dong,
  %``An Area-Law Prescription for Holographic Renyi Entropies,''
  arXiv:1601.06788 [hep-th].
  %%CITATION = ARXIV:1601.06788;%%
  %5 citations counted in INSPIRE as of 17 Jul 2016
  
%\cite{Nakaguchi:2016zqi}
\bibitem{Nakaguchi:2016zqi} 
  Y.~Nakaguchi and T.~Nishioka,
  %``A Holographic Proof of R\'enyi Entropic Inequalities,''
  arXiv:1606.08443 [hep-th].
  %%CITATION = ARXIV:1606.08443;%%

\end{thebibliography}
\end{document}